# On the Molecular Origin of the Cooperative Coil-to-globule Transition of Poly(N-isopropylacrylamide) in Water

L. Tavagnacco,[a,b*] E. Zaccarelli[a,b] and E. Chiessi[c,a*]

By means of atomistic molecular dynamics simulations we investigate the behaviour of poly(N-isopropylacrylamide), PNIPAM, in water at temperatures below and above the lower critical solution temperature (LCST), including the undercooled regime. The transition between water soluble and insoluble states at the LCST is described as a cooperative process involving an intramolecular coil-to-globule transition preceding the aggregation of chains and the polymer precipitation. In this work we investigate the molecular origin of such cooperativity and the evolution of the hydration pattern in the undercooled polymer solution. The solution behaviour of an atactic 30-mer at high dilution is studied in the temperature interval from 243 to 323 K with a favourable comparison to available experimental data. In the PNIPAM water soluble states we detect a correlation between polymer segmental dynamics and diffusion motion of bound water, occurring with the same activation energy. Simulation results show that below the coil-to-globule transition temperature PNIPAM is surrounded by a network of hydrogen bonded water molecules and that the cooperativity arises from the structuring of water clusters in proximity to hydrophobic groups. Differently, the perturbation of the hydrogen bond pattern involving water and amide groups occurs above the transition temperature. Altogether these findings reveal that even above the LCST PNIPAM remains largely hydrated and that the coil-to-globule transition is related with a significant rearrangement of the solvent in proximity of the surface of the polymer. The comparison between the hydrogen bonding of water in the surrounding of PNIPAM isopropyl groups and in bulk displays a decreased structuring of solvent at the hydrophobic polymer-water interface across the transition temperature, as expected because of the topological extension along the chain of such groups. No evidence of an upper critical solution temperature behaviour, postulated in theoretical and thermodynamics studies of PNIPAM aqueous solution, is observed in the low temperature domain.

## Introduction

The phase behaviour and self-assembly of macromolecules in aqueous solution are strongly affected by the hydration pattern, especially in case of amphiphilic polymers. For such systems the distribution of water molecules in the surrounding of chains, driven by the combination of polar, hydrogen bond and hydrophobic interactions, is modulated in a complex way by pressure and temperature, generating phase diagrams with differently shaped mixing gaps.[1, 2] Moreover, the sensitivity of the hydration shell on the state variables plays a major role in stabilizing specific chain conformations and can trigger temperature-dependent intramolecular and intermolecular association processes, such as the folding of proteins[3] and the protein aggregation leading to amyloid fibril formation,[4] respectively. An example of the influence of the hydration microstructure on the aqueous solution behaviour of amphiphilic macromolecules is provided by three polymers: poly(L-leucine), PL, poly(N-vinylisobutyramide), PNVIBA and poly(N-isopropylacrylamide), PNIPAM. In PL, PNVIBA, and PNIPAM the repeating unit has the same chemical composition and contains an amide and an isopropyl group, i.e. a hydrophilic and a hydrophobic moiety, respectively. For all polymers the hydrophobic group is located in the side chain. The amide functionality is inserted into the backbone of the polypeptide chain of PL whilst it is located in the side chain of the vinyl polymers. Such structural difference, which also affects the hydration shell of the macromolecules, has a huge effect on the water affinity. While PL is scarcely soluble,[5] PNVIBA and PNIPAM are water soluble at temperatures below a lower critical solution temperature, LCST.[6, 7] The structural difference between the two vinyl polymers is much more subtle, as compared to PL, consisting in an inversion of sequence of NH and CO groups within the side chain. However, it is sufficient to significantly vary the LCST value, being about 32 and 39 °C for PNIPAM and PNVIBA, respectively, for the same solution concentration and comparable degree of polymerization, DP.[8] By considering that the conformational characteristics of PNIPAM and PNVIBA are identical, this effect is totally ascribable to variations in the hydration pattern of the chains.

The phase behaviour of PNIPAM in aqueous solution motivated half century of research activity[9] and in the last decade this polymer has been considered as a pivotal component to realize stimuli responsive soft microdevices,[10] by translating the LCST behaviour of the linear chain into the temperature-induced volume phase transition, VPT, of PNIPAM based networks. Several investigations aimed at tuning the value of the LCST according to the requirements of specific applications. Variations of the aqueous medium, such as ionic strength changes,[11, 12] co-solvents[13-15] and co-solutes [16, 17] addition and of the polymer, as the insertion of heteroresidues in the chain scaffold and the modulation of the stereochemistry,[18, 19] allow to obtain an interval of LCSTs ranging about 40 K. In all these cases, an alteration of the organization of water molecules in the surrounding of the polymer can be invoked to explain the different phase transition behaviour of PNIPAM.

In 2005 Okada and Tanaka theoretically postulated a correlation between the flat behaviour of the phase separation line in the phase diagram of PNIPAM aqueous solution and the structure of the hydration shell, able also to explain other characteristics of the soluble-insoluble state transition of PNIPAM, such as the scarce dependency on the chain DP.[20] Their hypothesis is the formation of sequential hydrogen bonds involving polymer and water molecules in the surrounding of the PNIPAM chains, which is defined as a cooperative hydration. This model in fact allows to reproduce peculiar features of the phase diagram, highlighting that the precursor step of the soluble-insoluble state transition is an intramolecular process.[21] The picture of the cooperative hydration is, thereafter, often cited in the discussion of the behaviour of PNIPAM based systems in aqueous environment.[22-29] Interpretations at the molecular level of such cooperativity, based on indirect experimental findings, have been proposed. In the study of PNIPAM aqueous solutions by high-frequency dielectric relaxation techniques[30] it was raised the hypothesis that water molecules, directly hydrating the polymer, are involved in a network of HBs stabilizing the solvation shell. In particular, the formation of HB bridges between adjacent isopropylamide groups is supposed to play an important role in sustaining an extended conformation of PNIPAM below the LCST.[30] The entropic penalty caused by the hydrogen bonding of water in the surrounding of hydrophobic methyl groups, and its increasing contribution at increasing temperature, is postulated to be a determining factor for the cooperative collapse in an UV Resonance Raman investigation of PNIPAM nanogels.[24]

[a] CNR-ISC, Uos Sapienza, Piazzale A. Moro 2, 00185 Roma, Italy
[b] Dipartimento di Fisica, Sapienza Universita di Roma, Piazzale A. Moro 2, 00185 Roma, Italy. E-mail: letizia.tavagnacco@roma1.infn.it
[c] Department of Chemical Sciences and Technologies, University of Rome Tor Vergata, Via della Ricerca Scientifica I, 00133 Roma, Italy. E-mail: ester.chiessi@uniroma2.it

The thermodynamic approach of Okada and Tanaka, in addition to the description of the LCST behaviour of PNIPAM, predicts a "classical" upper critical solution temperature, UCST, phase separation, occurring in a temperature range below the water melting temperature.[20] An analogous UCST de-mixing gap for PNIPAM aqueous solutions is detected in the binodal and spinodal curves calculated by the phenomenological free energies of Afroze et al.[9, 31] The predicted value of the UCST is 15°C for a chain with DP=100,[9] a result hitherto experimentally unconfirmed. The occurrence of an UCST in the phase diagram of aqueous PNIPAM solution would imply a temperature/concentration regime where the Flory-Huggins parameter $\chi(T)$ decreases with increasing T, namely where the mixing entropy, $\Delta S_{mix}$, is positive. Such condition contradicts the typical solvation conditions of PNIPAM in water, with a favourable enthalpy contribution to mixing for the water interactions with amide moieties, but an unfavourable entropy contribution for hydrophobic effects.[7] Therefore the presence of an UCST should be correlated to a drastic change of hydration modality.

With such background, this study aims to explore the hydration pattern of PNIPAM across the LCST transition, in a temperature interval including the undercooled aqueous solution regime, by means of atomistic molecular dynamics, MD, simulations. Unlike previous MD simulations studies,[32-35] the stereochemistry of the polymer model is here designed for a realistic representation of atactic PNIPAM, which is the component of almost the totality of experimental PNIPAM based systems. The simulations reproduce the coil-to-globule PNIPAM transition according to the available experimental information and the results obtained from the trajectories analysis favourably compare with experimental data concerning both the hydration features and local polymer dynamics across the transition. With reference to the work of Okada and Tanaka and to the molecular mechanisms proposed in the literature for the coil-to-globule PNIPAM transition,[24, 30, 36] we analyse the markers of the cooperative hydration and show for the first time a differentiated thermal response of the hydration shell in the surrounding of hydrophobic and hydrophilic PNIPAM groups. In the temperature interval between 263 and 293 K we detect a correlation between polymer segmental dynamics and diffusion motion of water molecules associated to PNIPAM, that consolidates the picture of a hydration dependent behaviour for both the thermodynamics and the dynamics of this polymer. With our simulation model no discontinuity of the hydration modality of PNIPAM is observed in the undercooled and low temperature solution, that seems to exclude an UCST behaviour in water in the low temperatures domain.

## Computing details

### Model and simulations

The model consists of a linear polymer chain of 30 residues in water at infinite dilution. Such DP is equivalent to about one-two Kuhn segments of atactic PNIPAM in the water soluble state, according to the estimate of the characteristic ratio of high molecular weight single chains.[21, 37] Turbidity measurements as a function of temperature on aqueous solutions of the atactic PNIPAM oligomer composed of 28 repeating units display a behaviour similar to that observed for the atactic polymer with a DP of 100,[38, 39] for comparable PNIPAM concentrations. Moreover, previous MD simulation studies of syndiotactic and isotactic-rich PNIPAM 30-mers show that the single chain coil-to-globule transition can be detected with this DP.[32, 33]

The stereochemistry of the 30-mer was designed by assuming a Bernoullian distribution of meso and racemo dyads, for a final content of racemo dyad equal to 55 %. This dyad composition coincides with that of PNIPAM synthesized without stereo-selective agents[40, 41] and the stereoisomer can be assumed as a model of atactic PNIPAM. It is noteworthy that the tacticity has a strong influence on solubility and phase behaviour of this polymer in water, with an increase or a decrease of hydrophobicity with the increase of the degree of isotacticity or sindiotacticity, respectively.[42-47] The chain conformation of the initial structure was obtained by imposing to the backbone dihedral angles values corresponding to states of minimum conformational energy for the dyads composing the oligomer.[48, 49] The amide groups in side chain have a *trans* arrangement. The details on stereochemistry and the sequence of backbone conformational states of the initial structure are reported in Table S1 and Fig. S1 of the Electronic Supplementary Information (ESI).

The polymer was described by the force field OPLS-AA[50] with the modifications of Siu et al.,[51] successfully used to model the solution behaviour of the syndiotactic 30-mer.[52] The TIP4P/Ice model was selected for water,[53] to properly represent the undercooled aqueous solution. This water model predicts a melting temperature almost coincident with the experimental value, making comparable the simulation and experimental temperature conditions. Moreover, it provides a satisfactory description of liquid water.[53]

The polymer chain was centered in a cubic box of 9 nm side and oriented along a box diagonal to maximize the distance between periodic images. A minimization of energy in vacuum with tolerance of 10 kJ·mol⁻¹·nm⁻¹ was carried out. Then about 23000 TIP4P/Ice water molecules were added and a further energy minimization was performed with tolerance of 100 kJ·mol⁻¹·nm⁻¹. The resulting system was used as initial configuration of the simulations for 9 temperature conditions, namely from 243 to 323 K, every 10 K.

MD simulations were carried out in the NPT ensemble for 300 ns at each temperature. Trajectories were acquired with the leapfrog integration algorithm[54] with a time step of 2 fs; cubic periodic boundary conditions and minimum image convention were applied. The length of bonds involving H atoms was constrained by the LINCS procedure.[55] The velocity rescaling thermostat coupling algorithm, with a time constant of 0.1 ps[56] was used to control the temperature. The pressure of 1 atm was maintained by the Parrinello–Rahman approach, with a time constant of 2 ps.[57, 58] The cutoff of nonbonded interactions was set to 1 nm and electrostatic interactions were calculated by the smooth particle-mesh Ewald method.[59] The final 75 ns of trajectory were considered for analysis, sampling 1 frame every 5 ps.

The trajectory acquisition and analyses were carried out with the GROMACS software package (version 5.1.2).[60, 61] The graphic visualization was obtained by the molecular viewer software package VMD.[62]

**Trajectory analysis**

The solvent accessible surface area, sasa, of a solute molecule is the surface of closest approach of the centres of solvent molecules where both solute and solvent are represented by hard spheres. Computationally, this surface is defined as the van der Waals envelope of the solute molecule expanded by the radius of the solvent sphere about each solute atom centre.[63] In this work the PNIPAM sasa was evaluated using a spherical probe with radius of 0.14 nm and the values of Van der Waals radii of the work of Bondi.[64, 65] The distributions of sasa values were calculated with a bin of 0.1 nm².

The root mean square fluctuation, RMSF, of each backbone carbon atom was estimated as:

$$RMSF_i = \sqrt{\langle ([\vec{r_i} - \langle \vec{r_i} \rangle]^2) \rangle} \quad (1)$$

where $r_i$ and $\langle r_i \rangle$ are the instantaneous and the time averaged position of the backbone atom $i$, respectively. The RMSF was calculated within the time interval of 100 ps and over the whole production run. A roto-translational fit to the structure at the starting time was applied.

To analyse the torsional dynamics of the polymer backbone we monitored the transitions between trans and gauche states of the dihedral angles formed by backbone carbon atoms. The average lifetime of a rotational state $\langle \tau \rangle$ was estimated according to eq 2:

$$\langle \tau \rangle = \frac{t_{TOT} \cdot N_{DIHE}}{N_{TRANS}} \quad (2)$$

where $t_{TOT}$ is the investigated time interval, $N_{TRANS}$ is the number of transitions of the dihedrals and $N_{DIHE}$ is the total number backbone dihedral angles.

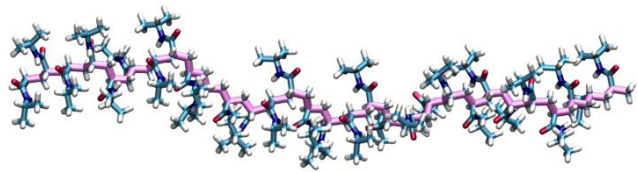

**Fig. 1** Starting conformation of atactic PNIPAM 30-mer.

An analogous analysis was carried out for the torsional dynamics of methyl groups in PNIPAM side chains, by considering the transitions of the dihedral angle $\phi$ formed by atoms N-C$_{isopropyl}$-C$_{methyl}$-H. The distribution of $\phi$ values was calculated at each temperature.

Hydration features were explored by the analysis of PNIPAM-water hydrogen bonds, HBs, and water-water HBs in the surrounding of the polymer. The occurrence of this interaction was evaluated adopting the geometric criteria of an acceptor−donor distance (A•••D) lower than 0.35 nm and an angle Θ (A•••D-H) lower than 60°, irrespective of the AD pair.

The ensemble of water molecules of the first hydration shell, FHS, was sampled by selecting molecules having the oxygen atom, OW, at a distance from nitrogen, N, or oxygen, O, or methyl carbon atoms of PNIPAM, lower than the first minimum distance of the corresponding radial distribution functions with OW atoms. Such distances are 0.35 nm for FHS of N and O atoms, and 0.54 nm for FHS of methyl carbon atoms of isopropyl groups. A water molecule was considered in the FHS of PNIPAM when it is within the FHS of at least one of these PNIPAM atoms.

At each time frame the hydrogen bonding between water molecules of the FHS was analysed. The distribution of values of the number of HBs per water molecule, N$_{wwHB,FHS}$, was calculated with a bin of 0.01.

The hydrogen bonding of bulk water was evaluated by sampling water molecules at a distance larger than 2 nm from the polymer.

The clustering of water molecules belonging to the first hydration of N and O atoms was analysed by calculating the population of clusters formed by hydrogen bonded solvent molecules.

The polymer influence on the solvent dynamics was investigated by considering the diffusion coefficient, D, of water molecules associated to PNIPAM. The D value was obtained from the long-time slope of the mean square displacement:

$$D = \frac{1}{6} \lim_{t \to \infty} \frac{d}{dt} \langle |r(t) - r(0)|^2 \rangle \quad (3)$$

where r(t) and r(0) correspond to the position vector of OW atom at time t and 0, respectively, with an average performed over both time origins and water molecules. To evaluate the limiting slope, a time window of 1 ns was considered, including in the ensemble average only the water molecules residing in the FHS at the beginning and at the end of the time interval. The population of this water ensemble consists, on average, of about 300 and 80 molecules at the lowest and highest temperature, respectively. The diffusion coefficient of bulk water, $D_0$, was estimated according to eq (3) by sampling water molecules at a distance larger than 2 nm from the polymer.

## Results and discussion

We first present results concerning polymer properties and compare them with experimental information on the coil-to-globule PNIPAM transition. Then we report the analysis of the polymer-water interactions and of the structural and dynamical features of water in the surrounding of the polymer. These data will be discussed in the light of the cooperative hydration model.

### Polymer properties

*Radius of gyration and solvent accessible surface area.* MD simulations were carried out in the temperature range between 243 K and 323 K, namely from the undercooled solution up to above the LCST value of the atactic polymer, typically of 304-306 K.[9] The coil-to-globule transition is the intramolecular process that precedes the chains aggregation and the precipitation of the polymer at the LCST. For high DP single chains of PNIPAM in water the coil-to-globule transition temperature, $T_{C \to G}$, is about 305 K.[21] Atactic PNIPAM oligomers with a chain length similar to that of this model display the same soluble-insoluble state transition in water at increasing temperature, as observed for the polymer.[38, 39] For such low DP chains the value of the LCST detected by turbidity experiments, representing the temperature condition where chains aggregation occurs, is very sensitive on PNIPAM concentration. LCST increases at increasing dilution and a value of about 313 K was observed for the 28-mer at the lowest investigated polymer concentration of 0.05 % (w/w).[38] Experimental data on the value of $T_{C \to G}$ for PNIPAM oligomers with a DP of about 30 are not available. Concerning the solution behaviour of PNIPAM in the low temperature regime, a glass transition temperature, $T_g$, of about 223 K is reported for 50 % (w/w) PNIPAM aqueous solutions and

lower $T_g$ values can be extrapolated at lower polymer concentrations.[31]

All simulations were performed with the same PNIPAM starting configuration, characterized by an extended geometry, as displayed in Fig. 1. The procedure allowed us to evaluate the temperature behaviour of the PNIPAM chain without any influence of the system history at a different temperature. In addition, being the initial structure in a state of minimum conformational energy, its evolution in the simulation highlights the influence of interactions with water.

In order to detect the coil-to-globule transition and the corresponding variation of water affinity of the polymer, the radius of gyration $s$ and the solvent accessible surface area, defined in section Trajectory Analysis, were monitored. The time behaviour of $s$ as a function of temperature is displayed in Fig. 2a and the distributions of sasa values in the production run are reported in Fig. 2b. The inspection of Fig. 2a leads to identify three temperature regimes: i) an undercooled solution regime, including the temperatures of 243 and 253 K, where only extended chains conformations are populated ($s$ values always larger than 1.6 nm); ii) a temperature interval including water soluble states of PNIPAM, from 263 to 293 K, where rearrangements of the coil conformation are present; iii) a temperature interval corresponding to PNIPAM water insoluble states, from 303 to 323 K, where the globule conformation is sampled. In particular, at 303 K the simulation detects a decrease of the polymer size in the first 50 ns trajectory, bringing to a stable chain conformation with very small $s$ fluctuations. This behaviour is ascribable to the single chain coil-to-globule transition of the oligomer, that occurs at a transition temperature comparable with that experimentally determined for the polymer.[21, 66] The chain conformation in the final snapshot at 303 K is illustrated in Fig. 2c. At the temperatures above 303 K, the coil-to-globule transition is detected after a longer trajectory interval, approximately of 170 ns. The time behaviour of $s$ at 313 K, Fig. 2a, shows an event of swelling-recoiling of the polymer chain in the 220-260 ns range, which is subsequent to the coil-to-globule transition. A similar reversibility of the coil-to-globule transition was detected in a previous MD simulations study of the isotactic 30-mer.[34] We suggest that these events may favour the interchain association, occurring at 313 K for highly dilute solutions of the atactic 30-mer.[38]

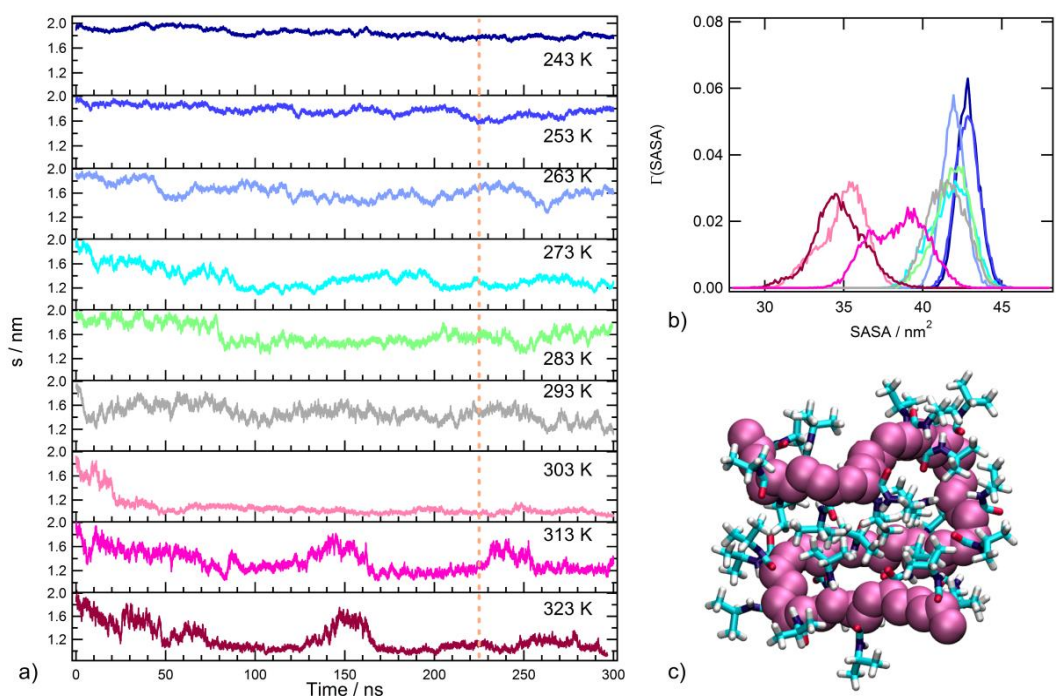

**Fig. 2** a) Time evolution of the radius of gyration (s) of PNIPAM chain as a function of the temperature. Dashed lines indicate the time interval considered for data analysis. b) Distributions of values of solvent accessible surface area as a function of the temperature (243 K, 253 K, 263 K, 273 K, 283 K, 293 K, 303 K, 313 K, 323 K are shown in dark blue, blue, light blue, cyan, green, yellow, pink, magenta and violet, respectively). c) Snapshot from the MD simulation of the PNIPAM chain at 303 K showing the collapsed conformation.

The behaviour of the solvent accessible surface area, displayed in Fig. 2b, summarizes the trend of *s* behaviour with temperature. The drastic decrease of sasa, observed at the temperatures above 293 K, identifies the transition between water soluble and insoluble states. Moreover, Fig. 2b allows us to distinguish the undercooled solution regime, characterized by higher sasa values. The backbone conformational transition observed at 303 K is concerted with a rearrangement of side chain groups to minimize the exposed hydrophobic surface area, as expected for PNIPAM hydrophobic collapse. The main contribution to the clustering of hydrophobic groups in the globule state comes from contacts between i, i + 2 repeating units and between more topologically distant residues (Fig. S2 of ESI). Overall, the temperature dependence of these structural parameters shows that our simulation setup, using the tip4p/ICE water model, is able to reproduce the PNIPAM coil-to-globule transition in a temperature range corresponding to the experimental one.

*PNIPAM-PNIPAM hydrogen bonding.* The correlation between chain conformation and intramolecular interactions is illustrated by the analysis of the hydrogen bonding between PNIPAM residues as a function of temperature. The temperature dependence of the average number of intramolecular hydrogen bonds per PNIPAM residue is showed in Table 1. On average, one inter-residue HB every 17 repeating units is detected at temperatures below the transition temperature and approximately a doubling of these interactions is observed at temperatures above 303 K.

**Table 1** PNIPAM-PNIPAM hydrogen bonding.

| Temperature (K) | PNIPAM – PNIPAM hydrogen bonds (*per* PNIPAM residue)* |
|---|---|
| 243 | 0.07 ± 0.01 |
| 253 | 0.05 ± 0.01 |
| 263 | 0.06 ± 0.01 |
| 273 | 0.04 ± 0.01 |
| 283 | 0.06 ± 0.01 |
| 293 | 0.07 ± 0.01 |
| 303 | 0.10 ± 0.01 |
| 313 | 0.12 ± 0.02 |
| 323 | 0.12 ± 0.02 |

*\* Errors are estimated by applying the blocking method.*

The coil-to-globule transition of PNIPAM and the subsequent intermolecular association in aqueous solution were studied by Fourier transform infrared spectroscopy.[67] Experimental data show an increase of intramolecular hydrogen bonding across the transition temperature and suggest that about 13% of the CO groups forms intra- or interchain hydrogen bonding in the globule state. Simulations agree with these findings and the value of 0.12 inter-residue HBs per PNIPAM repeating unit, obtained above the transition temperature (Table 1) is in very good agreement with the experimental estimate. Therefore, even in the collapsed state, PNIPAM side chains are not predominantly involved in interamide interactions, as highlighted in an UV Resonance Raman scattering study of PNIPAM nanogels and solutions, showing the prevalent hydrogen bonding of amide groups with water even above the VPTT and the LCST.[24]

*PNIPAM segmental and side chain dynamics.* We further explored the influence of the conformational rearrangement associated to the coil–to-globule transition on the segmental and side chain mobility of PNIPAM. Fig. 3 displays the root mean square fluctuation of the backbone carbon atoms as a function of temperature, with a time average over the production run. The higher mobility of chain ends as compared to the internal region of the chain is a temperature independent feature, as expected because of the lower topological constraint. Excluding the first and last five residues of the polymer, in the undercooled solution the backbone fluctuations have low widths (comparable to the C-C bond length) and are uniform along the chain. Above 253 K the segmental dynamics is activated and a non-homogeneous mobility is detected for the internal residues. The transition temperature is characterized by a significant mobility drop, since the extent of backbone fluctuations at 303 K is similar to that detected at 253 K. Therefore, the coil-to-globule conformation transition induces a quenching of the local dynamics which is equivalent to a 50 K temperature decrease. This simulation result agrees with the strong attenuation of segmental mobility observed for PNIPAM at the LCST by means of time-resolved fluorescence anisotropy measurements.[68] Fig. 3 shows that above 303 K the local mobility increases again, which is consistent with the observation of collapse-swelling events at 313 K and the higher fluctuation of the radius of gyration at 323 Ks, visible in Fig. 2a.

The local vibrational dynamics of PNIPAM in aqueous solution at a polymer concentration of 27 % (w/w) was investigated by means of quasi-elastic neutron scattering, QENS. The mean squared amplitude of these motions at temperatures between 283 and 323 K was estimated for atactic PNIPAM with a DP of about 290. The experiments detect a decrease of mobility at 303 K, persisting even at the higher temperatures.[69] To compare the simulation findings with this information, we calculated the root mean square fluctuation of the backbone carbon atoms in the time window of 100 ps, corresponding to the QENS measurement time resolution. The average value of the fluctuation, calculated not including the first and last five residues of the chain, is reported in the inset of Fig. 3. Over the time window of one hundred ps the simulations show a monotonic increase of local mobility from 243 to 293 K and a drop at 303 K. The fluctuations at 283 and 293 K are consistent with the values reported in ref. 67 and the reduction of mobility detected in the simulation at 303 K is comparable to that measured across the LCST. Simulations display higher fluctuation values at 313 and 323 K, as compared to the value observed at 303 K. This discrepancy with the QENS result can be ascribed to the effect on the segmental mobility of chains aggregation, that occurs in the concentrated PNIPAM solution but it is not modelled in these simulations.

With the aim of analysing the conformational mobility of the polymeric chain we monitored the transitions between rotational states of the backbone dihedral angles along the trajectory. The values of the average lifetime of a rotational state $<\tau>$, estimated as described in the Trajectory Analysis section, are reported in Table S2 of ESI. In the temperature interval of thermodynamic stability of the polymer solution, the characteristic time of this local motion is of tens of nanoseconds. The average time scale for molecular motions of PNIPAM residues, including both main chain moieties and isopropylamide groups, was estimated by [13]C NMR relaxation measurements on aqueous ($D_2O$) solution at 298 K, for a PNIPAM concentration and DP of 20 % (w/w) and 2500, respectively.[70] The data of relaxation time T1 yielded a rotational relaxation time of about 30 ns for PNIPAM repeating units. The $<\tau>$ value obtained at 293 K is comparable to the experimental finding, by considering the differences between our model and the experimental system. The simulations detect an attenuation of torsional mobility at 303 K, as observed for the RMSFs of backbone carbon atoms. The $<\tau>$ values obtained below the coil-to-globule transition temperature follow the Arrhenius behaviour with activation energy, Ea, of 40±2 kJ mol$^{-1}$ (Fig. S3a of ESI).

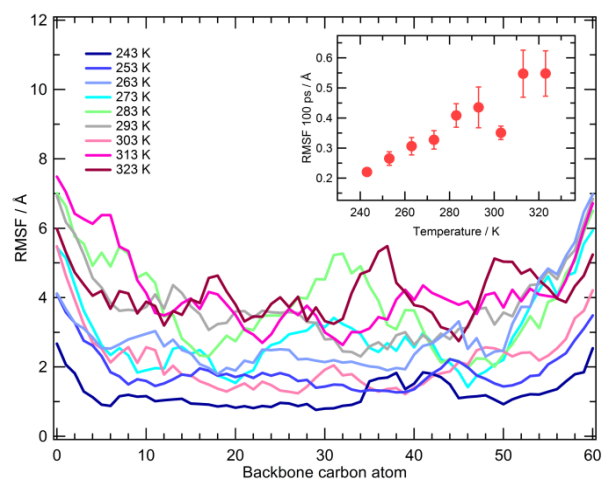

**Fig. 3** Root mean square fluctuations of PNIPAM backbone carbon atoms averaged over the last 75 ns of the simulations. Results are shown for different temperatures. The inset displays the RMSF averaged over 100 ps, not including the first and last five residues of the chain. Errors are estimated by applying the blocking method.

We carried out an analogous analysis for the torsional dynamics of methyl groups in PNIPAM side chains. The rotational lifetimes for this motion, reported in Table S2 of ESI, range from hundreds to tens of ps and an activation energy of 18 ± 2 kJ mol$^{-1}$ can be estimated (Fig. S3b of ESI). In this case no dynamic discontinuity was detected across the transition temperature. The methyl group dynamics in glassy polyisoprene was studied by neutron backscattering experiments. Between 2 and 120 K the methyl group rotational correlation time displays an Arrhenius-like behaviour with Ea ≈ 13 kJ mol$^{-1}$.[71] A Gaussian distribution of activation energies for methyl torsional transitions with mean around 15 kJ/mol and standard deviation around 3 kJ/mol was obtained for atactic polypropylene by molecular simulations and QENS experiments.[72] The average methyl rotation barrier in the hydrophobic core of proteins was estimated as equal to 12 ± 4 kJ mol$^{-1}$ from 2H relaxation data.[73] The activation energy obtained from simulations (Fig. S3b of ESI) is comparable with these experimental Ea values, taking into account the different structure and environment of our system. Fig. S4 of ESI displays the distribution of values of the methyl dihedral angle, $\phi$. In our simulations this local motion is fully activated also at the lowest temperature, accordingly to the behaviour observed for other polymers bearing side chain methyl groups.[71, 72] The torsion of side chain methyl groups is the first kind of motion to be switched on at low temperature in glassy polymers, followed, at much higher temperatures, by the backbone segmental dynamics.[71] Similarly, methyl group rotation is activated in proteins from about 100 K upwards, whist the motions involving jumps between minimum energy conformations occur at temperatures about one hundred K higher.[74-77] In agreement with these results, we find a sensibly higher activation energy for the backbone torsional dynamics, as compared to the $E_a$ associated to the methyl groups rotation.

**PNIPAM hydration**

*PNIPAM-water hydrogen bonding.* The water affinity of PNIPAM is quantified by the polymer-water hydrogen bonding. The fully hydrogen bonded state is characterized by 3 HBs *per* repeating unit, where 2 HBs are formed with the hydrogen acceptor CO group and 1 HB with the hydrogen bonding donor NH group. Fig. 4 displays the average number of hydrogen bonds between PNIPAM and water, normalized to the number of residues, as a function of temperature. Below the coil-to-globule transition temperature, more than 2.5 HBs *per* PNIPAM residue are formed with water, indicating that the amide functionalities are predominantly fully hydrated. Such finding agrees with the experimental estimate, indeed results of spectroscopy measurements indicate that below the LCST each amide group of PNIPAM is hydrated by two (to three) water molecules as a result of hydrogen-bond formation.[78, 79] A similar evidence was found for PNIPAM nano-hydrogels below the VPTT.[24] Our simulations detect a reduction of PNIPAM-water HBs across the transition temperature, and the sigmoidal profile showed in Fig. 4 is reminiscent of the behaviour predicted by Okada and Tanaka for the PNIPAM bound water in the soluble-insoluble state transition.[20] The decrease of polymer-water HBs finds a correspondence with the increase of intramolecular HBs (Table 1). The molecular-level structural changes of PNIPAM in aqueous solution were explored by a Compton scattering study, highlighting a reduction of the amount of HBs formed by polymer with water after the coil-to-globule transition, however the change in the number of hydrogen bonds was not quantified.[80] A reduction of about one polymer-water HB *per* PNIPAM residue was estimated in the volume phase transition of PNIPAM nanogels by UV Resonance Raman scattering experiments.[24] As compared to the last result, the decrease of PNIPAM-water HBs detected in the simulations across the coil-to-globule transition temperature, about 0.5 HB per residue, is lower. The discrepancy can be explained by considering that our system models the single chain transition whilst the rearrangement in the chemically cross-linked polymer network of the nanogel, occurring at the volume phase transition, implies both intra- and interchain aggregation. In these conditions, a higher extent of dehydration, as compared to the polymer at infinite dilution, is expected. Simulation results in Fig. 4 highlight a characteristic of PNIPAM which is often misinterpreted in the literature.[81] Even in the globule state amide groups are considerably hydrated, therefore PNIPAM is not a hydrophobic material.

The main physical factor determining the occurrence of the coil-to-globule transition at $T_{c \to g}$ is the concomitant decrease of the excluded volume to water, that leads to an increase of the translational entropy of solvent molecules. The contribution of this excluded volume effect raises at increasing temperature in water, since water density is scarcely affected by temperature because of the HB strength, up to make the transition thermodynamically possible. The same argument explains the cold denaturation observed in intrinsically disordered proteins and in globular proteins under special pH conditions or in the presence of denaturant.[82] A different theory, which is also focused on the excluded volume effect and the translational entropy of water but accounts for the essential role of hydrogen bonds of water in the temperature dependence of the excluded volume effect, has been presented.[83] This theory is capable of elucidating the cold denaturation of yeast frataxin occurring at 280 K under physiological condition.[84] In this respect, the analogy between the PNIPAM coil-to-globule transition and the protein cold denaturation was already highlighted in the literature.[85] However, the abrupt variation from a water soluble to a water insoluble state at the coil-to-globule transition temperature suggests a relevant rearrangement of the solvent in proximity of the polymer. These structural modifications can contribute to the entropy and enthalpy variations at the transition and modulate the value of the LCST.[82] An example of this effect is found in the difference between PNVIBA and PNIPAM LCSTs.[6, 7] The structure of water in the surrounding of polymers was shown to depend on the polar or nonpolar character of side chains.[79] Therefore, changes in the water ensemble enveloping the polymer before and after the coil-to-globule transition can be expected because of the variation of the hydrophilic/hydrophobic surface ratio in the solvent accessible surface of the chain. To detect the temperature dependent modifications of structural and dynamical features of water directly interacting with PNIPAM is the main aim of our investigation.

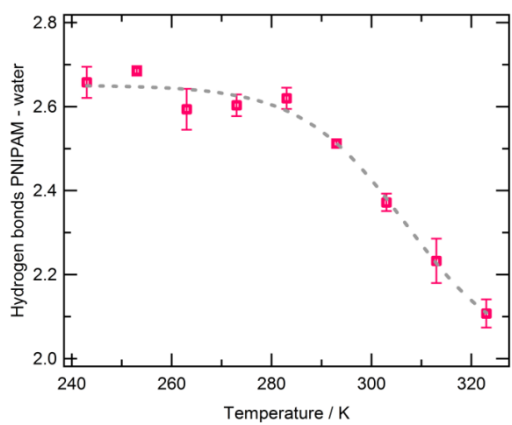

**Fig. 4** Average number of hydrogen bonds between PNIPAM and water *per* PNIPAM residue as a function of the temperature. Errors are estimated by applying the blocking method.

For this analysis we use an operative definition of hydration water based on a distance criterion, by selecting the solvent molecules belonging to the first hydration shell, FHS, of both hydrophilic and hydrophobic side chain atoms (see Trajectory analysis section). With such a definition, an average number of hydration water molecules *per* PNIPAM residue of 14 ± 1 is calculated in the temperature range preceding the transition (see Table S3 of the ESI). Dielectric relaxation measurements on PNIPAM aqueous solutions indicate that the hydration number of this polymer below the LCST is about 11 or 13,[70, 86] in satisfactory agreement with the estimate from simulations.

*Hydrophilic and hydrophobic hydration.* It is acknowledged in the literature that the coil-to-globule transition occurs with a concomitant de-hydration of the polymer, however the precise number of hydrated water molecules in phase-separated PNIPAM chains above the LCST is debated due to the difficulty of experiments.[30] This de-hydration is observed in the simulations, as showed by the drop of the number of FHS water molecules at 303 K (Table S3 of the ESI). Fig. S5 of the ESI displays the temperature dependence of the number of water molecules belonging to the first hydration shell of N and O atoms or to the first hydration shell of carbon atoms of methyl groups. This plot reveals that the main contribution to the PNIPAM de-hydration comes from the loss of water in proximity of hydrophobic groups. A recent neutron scattering investigation on 25 % (w/w) PNIPAM solutions describes the evolution of hydration water across the LCST.[27] Interestingly, a splitting in two populations is observed at the transition temperature for water molecules associated to the polymer, one of these populations corresponding to a very strongly bounded water. Overall, the decrease of hydration water molecules, comparing water soluble and insoluble states, amounts to about 14 %,[27] the same value detected in our simulations (Table S3 of the ESI).

In the experimental study of ref.[30] Ono and Shikata raised the hypothesis that water molecules, directly hydrating the polymer, are involved in a network of HBs stabilizing the solvation shell. In particular, the formation of HB bridges between contiguous isopropylamide groups was supposed to play an important role in maintaining an extended conformation of PNIPAM below the LCST. According to this description, the cooperative loss of the hydrogen bond bridges at the LCST represents an essential element for the phase transition of the aqueous PNIPAM solution. A cooperative hydration pattern of the polymer was similarly postulated by Okada and Tanaka, which theoretically derived the phase diagram of PNIPAM in water.[20] However, their interpretation is mainly focussed on the cooperativity in the hydration of hydrophilic moieties. To compare with these hypotheses the results of our simulations, we analysed the HB connectivity of the ensemble of water molecules composing the first hydration shell. The distribution of values of the number of HBs formed between water molecules belonging to the PNIPAM FHS, normalized *per* water molecule, $N_{wwHB,FHS}$, is shown in Fig. 5. Monomodal distributions are obtained at the temperatures of 243 and 283 K, exemplifying the conditions of undercooled and thermodynamically stable solution, respectively, with maximum at the same $N_{wwHB,fhs}$ value of about 1.42 HBs per molecule. At 303 and 323 K, the distributions enlarge and are translated toward lower abscissa values. This behaviour denotes a reduction of connectivity between FHS water molecules after the coil-to-globule transition, which is consistent with the pictures proposed by Ono and Shikata[30, 70] and by Okada and Tanaka.[20]

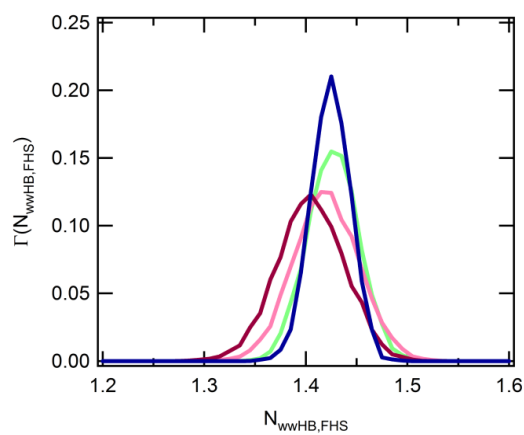

**Fig. 5** Distribution of values of the number of water-water HBs within the PNIPAM FHS at 243 K (dark blue), 283 K (green), 303 K (pink) and 323 K (violet). Abscissa values are normalized *per* water molecule.

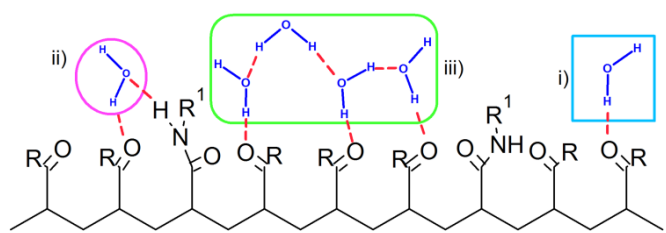

**Scheme 1** Schematic representation of the hydrophilic PNIPAM-water interface. Only hydrogen bonded groups of PNIPAM are explicitly shown, R and R1 being –NH-CH(CH3)2 and -CH(CH3)2, respectively. Water molecules are hydrogen bonded to the polymer with one of these three modalities: i) as a single water molecule (blue square); ii) as a water molecule simultaneously hydrogen bonded to two amide groups (pink circle); iii) as a water molecule hydrogen bonded to other water molecules of the hydrophilic FHS (green rectangle).

The results in Fig. 5 suggest a dependence of the hydrogen bonding in the FHS water on PNIPAM backbone conformation even beyond the temperature induced coil-to-globule transition. By analysing the simulation trajectory at 293 K we find a correlation between hydrogen bonding of FHS water and end-to-end distance, consisting in a weak increase of HBs number between water molecules with the increase of chain elongation (Fig. S6a of the ESI). On the contrary, the polymer-water hydrogen bonding is unaffected (Fig. S6b of the ESI). This finding favours a comparison with the results of an Atomic Force Microscopy, AFM, investigation of atactic single chain PNIPAM in aqueous solution at room temperature.[87] In this work a free energy increase of 5 kJ *per* mol of PNIPAM residues was estimated for the rearrangement of hydration water in the transition from a relaxed coil to an elongated conformation. According to simulation results, the isothermal elongation originates an increase of water hydrogen bonding (Fig. S6a) with a corresponding negative enthalpy variation. Therefore the free energy cost estimated by AFM has to be ascribed to an unfavourable entropy contribution.

We performed a selective analysis of the hydration pattern of hydrophilic and hydrophobic PNIPAM moieties to identify temperature and conformation dependent discontinuities that can reveal the molecular mechanism of the PNIPAM de-hydration across the transition.

The water molecules within the FHS of PNIPAM nitrogen and oxygen atoms represents about the 16 % of FHS water. Almost the totality of these solvent molecules, at the hydrophilic interface of PNIPAM (Scheme 1), are hydrogen bonded to the polymer, typically with one of these three modalities: i) as a single water molecule; ii) as a water molecule simultaneously hydrogen bonded to two amide groups; iii) as a water molecule hydrogen bonded to other water molecules of the hydrophilic FHS. Fig. 6a displays the contribution of these three kinds of hydrophilic solvation as a function of temperature, with normalization *per* PNIPAM repeating unit. Such analysis highlights that the majority of the water molecules involved in HB with PNIPAM is inter-connected, being the type iii) predominant, and that more than one water molecules every three PNIPAM residues bridges two amide groups (type ii) ). The individually hydrogen bonded water molecules (type i) ) are about the 30 % of total hydrophilic shell below the coil-to-globule transition temperature. Three temperature dependent regimes of hydrophilic hydration can be identified in Fig. 6a, which are correlated to the temperature regimes previously defined by the behaviour of PNIPAM properties. In the temperature interval corresponding to the thermodynamically stable polymer solution, from 263 to 293 K, the simulations show that a HB scaffold formed by both PNIPAM-water and water-water hydrogen bonds envelops the polymer chain. The structural details of this scaffold are slightly different in the undercooled aqueous solution, at 243 and 253 K, since the interconnection between hydrophilic FHS water molecules and the water bridging are higher and lower, respectively. However these variations self-compensate, therefore a high HB connectivity in the surrounding of PNIPAM is detected below the transition temperature, even in undercooling condition. At 313 and 323 K, above the coil-to-globule transition temperature, both the water-water HBs and the bridged water molecules display a drop. This discontinuity in the hydration pattern of hydrophilic groups, observed between 303 and 313 K, denotes a decrease of connectivity within the hydrophilic domain of PNIPAM that can be associated to a molecular mechanism of cooperative de-hydration. However, such discontinuity is detected not at the transition temperature but above it. Only the population of water molecules individually hydrogen bonded to PNIPAM displays a variation at $T_{C \rightarrow G}$. To clarify this behaviour, we analysed the clustering of water molecules within the hydrophilic FHS as a function of temperature. Fig. 6b shows the distribution of the cluster sizes, expressed as number of water molecules interconnected by HB. It is noteworthy that the clusters of 2 and 3 water molecules are those most frequently observed, since their molecular size well fits the region between amide groups of near-neighbouring residues. The decrease of the number of such clusters across the transition temperature, visible in Fig. 6b is coherent with the polymer dehydration. However, only at 303 K few clusters with 7-8 water molecules are detected. This result can be explained by considering the rearrangement of the hydrophilic polymer-water interface that occurs during the coil-to-globule transition. In the compact globule (Fig. 2c) topologically distant amide groups, in addition to isopropyl

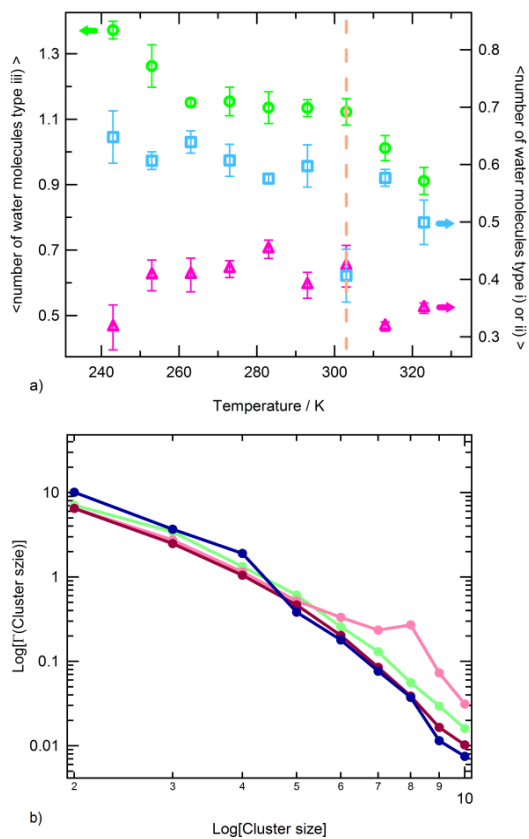

**Fig. 6** a) Temperature dependence of the average number of water molecules surrounding hydrophilic PNIPAM functionalities and hydrogen bonding the polymer: type i) individual water molecules (blue squares); type ii) water molecules simultaneously hydrogen bonded to two amide groups (pink triangles); type iii) water molecules hydrogen bonded to other water molecules of the hydrophilic FHS (green circles). Errors are estimated by applying the blocking method. The vertical dotted line indicates the transition temperature. b) Distribution of cluster sizes of water molecules hydrating hydrophilic regions of the PNIPAM chain at 243 K (dark blue), 283 K (green), 303 K (pink) and 323 K (violet).

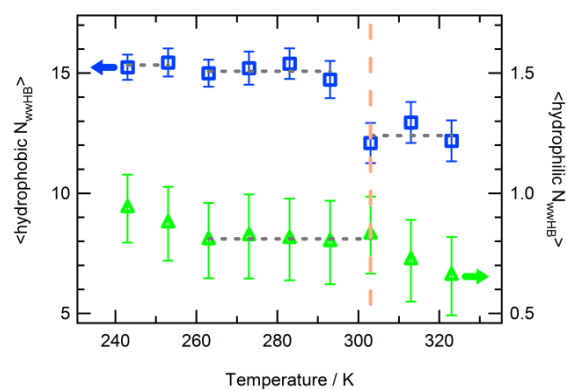

**Fig. 7** Temperature dependence of the average number of water-water hydrogen bonds *per* residue in the PNIPAM FHS. HBs formed between water molecules in the surrounding of hydrophobic and hydrophilic polymer groups are displayed with blue squares and green triangles, respectively. Data represent time averaged values and standard deviation.

groups, move close. In proximity to such hydrophilic pockets the formation of larger water cluster is favoured. Above $T_{C \rightarrow G}$ the higher conformational mobility of the chain (Fig. 2a) destabilizes water clusters with large size in the surrounding of PNIPAM hydrophilic groups.

The hydrogen bonding connectivity of the water molecules belonging to the hydrophilic and hydrophobic FHS of PNIPAM are compared in Fig. 7, in the investigated temperature interval. A pronounced discontinuity in the number of water-water HBs formed at the hydrophobic interface of PNIPAM is detected at 303 K, namely at the coil-to-globule transition. Differently, the HB discontinuity at the hydrophilic interface occurs at 313 K. These findings lead to identify the mechanism of the cooperative hydration, concerted to the conformational collapse of the chain, with the disruption of hydrogen bonds in the surrounding of isopropyl groups, according to the hypothesis of Ono and Shikata.[30, 70]

The results in Fig. 7 show that the perturbation of the HB network formed by water molecules bound to amide groups needs a higher temperature as compared to the coil-to-globule transition temperature. The thermal behaviour of PNIPAM in its concentrated $D_2O$ solution (20 wt %) was studied by FTIR and 2D-IR correlation spectroscopy.[36] In this work the microdynamics phase separation mechanism is described as consisting of the preliminary dehydration of the $CH_3$ groups, followed by the main-chain diffusion and aggregation, and finally the hydrogen bond transition of amide groups. By comparing this mechanism with the modality of PNIPAM de-hydration observed in the simulations, we can explain the late de-hydration of hydrophilic moieties with the absence of inter-chain aggregation. It is noteworthy that the temperature of 313 K, where we observe the rearrangement of hydrophilic hydration shell, corresponds to the temperature where the aggregation of the PNIPAM 30-mer is experimentally detected.

The enthalpy variation of the coil-to-globule transition, $\Delta H_{C\rightarrow G}$, in PNIPAM diluted aqueous solution was measured by differential scanning calorimetry,[15] obtaining a value of about 5.5 kJ/mol of repeating unit. The endothermal character of the transition is mainly attributed to the perturbation of the structure of water molecules in the surrounding of the polymer and to the release of water toward the solution bulk. This leads to a net reduction of structuring, namely of hydrogen bonds involving both polymer and water.[15] Simulation results indicate that the coil-to-globule transition occurs with the loss of about 0.5 PNIPAM-water HBs (Figure 4) and the gain of about 0.07 PNIPAM-PNIPAM (Table 1) per PNIPAM residue. Data in Fig. S5 of the ESI show that about 2 water molecules are expelled from the hydrophobic groups shell of a PNIPAM residue in the transition. To evaluate the variation of hydrogen bonding of these released molecules, we analysed the hydrogen bonding of a water ensemble corresponding to bulk water and compared with hydrogen bonding of water near hydrophobic groups. The results are reported in Table S4 of the SI.

The average number of HB per water molecules in bulk is about 2.2, in good agreement with the experimental value of 2.2 ± 0.5 at 303 K.[88] Moving from PNIPAM hydrophobic shell to solution bulk, water experiences an increase of water-water hydrogen bonding equal to

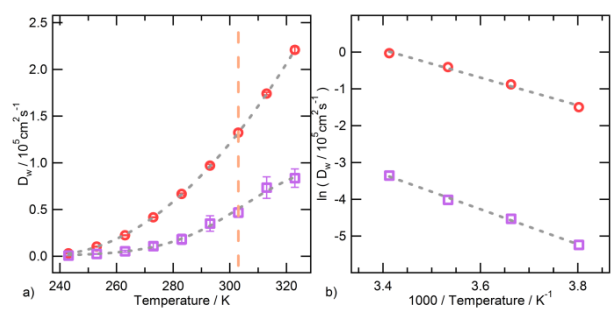

**Fig. 8** a) Temperature dependence of water translational diffusion coefficient as calculated for bulk water (red circles) and hydration water (violet squares). Dashed lines are guide to the eye. The vertical dotted line indicates the transition temperature. b) Arrhenius plot of water translational diffusion coefficient with the linear fit of the data. Errors are estimated by applying the blocking method.

about 0.8 HBs per water molecule. By considering that the enthalpies of formation of amide-water and amide-amide HB are -50.7[89] and -41 kJ/mol,[90] respectively, and the enthalpy of formation of water-water HB is -10.6 ± 0.5 kJ/mol,[91] an overall change in enthalpy of about 7 kJ/mol can be estimated for the transition, from the hydrogen bonding variations just described. This result favourably compares with the experimental value of $\Delta H_{C\rightarrow G}$ for PNIPAM in water.[15]

The comparison between the hydrogen bonding properties of water near PNIPAM hydrophobic groups and of bulk water, reported in Table S4, highlights a relevant aspect of PNIPAM hydration. At the hydrophobic polymer-water interface solvent molecules do not experience an increase of structuring, as compared to bulk. This feature is caused by the topological extension of the hydrophobic interface along the polymer chain, as stated by Chandler.[92] Such a result indicates that the driving force of the PNIPAM hydrophobic collapse at $T_{C\rightarrow G}$ does not derive from the rotational-entropy gain of water molecules released from the hydrophobic shell. Indeed, the contribution of a similar effect on the hydrophobic aggregation was shown to be minor around room temperature and decreasing at increasing temperature, as compared to the translational-entropy gain of all water molecules for the excluded volume decrease.[93, 94] The latter effect makes the transition thermodynamically favoured. However, our work demonstrates that the structuring of water in the surrounding of hydrophobic groups of PNIPAM, and in particular the connectivity between water molecules of the hydration shell, is responsible for the cooperativity of the coil-to-globule transition.

*Water dynamics.* We investigated the polymer influence on dynamics of water by comparing the diffusion coefficient of water molecules in the FHS, D, with that of bulk water, $D_0$. The temperature dependence of D and $D_0$ values, and the corresponding Arrhenius plots in the temperature interval where the polymer solution is thermodynamically stable, are illustrated in Fig. 8. An underestimate of $D_0$ values from simulations[95, 96] is typical for the tipe4p/ICE computational model.[97] The activation energy for the diffusion motion of bulk water (Fig. 8b) is equal to $E_a^{bulk}$ = 24 ± 1 kJ mol[-1], as compared to the experimental value of $E_a^{exp}$ = 19 kJ mol[-1].[98] The comparison between D and $D_0$ values, displayed in Fig. 8a, highlights a significant attenuation of water mobility in the surrounding of the polymer. Below the transition temperature the $D/D_0$ ratio ranges from 0.2 to 0.3 and it raises up to 0.4 above the transition temperature. QENS experiments carried out on PNIPAM aqueous solutions detected a similar slowdown of hydration water dynamics, with water diffusion coefficients less than half of the value for bulk water.[27] In addition, Fig. 8a shows a sigmoidal behaviour of the diffusion coefficient of hydration water as a function of temperature, differing from the monotonic increase observed for $D_0$ values. The inflection point can be identified at the transition temperature. These results are in agreement with a previous QENS study on 27 % (w/w) PNIPAM aqueous solution[69] which showed the presence of different regimes of hydration water dynamics in the temperature regions below and above the LCST. The activation energy obtained from the Arrhenius plot of D values below the transition temperature (Fig. 8b) is $E_a^{hydration}$ = 39 ± 2 kJ mol[-1] , sensibly higher as compared to the activation energy detected for the diffusion motion of bulk value. This finding conveys the mobility constraints of the water ensemble sampled with our analysis. Indeed the diffusion coefficient is calculated for water molecules persisting in the FHS for at least 1 ns and the majority of such molecules are hydrogen bonded to amide groups simultaneously forming the hydrogen bond network in the surrounding of the polymer. Dielectric relaxation experiments on PNIPAM aqueous solutions showed that the activation energy for the exchange or dehydration process of water molecules associated to the polymer is higher as compared to the activation energy related to the roto-translational dynamics of bulk water. The experimental estimate of this increase is equal to about 33 %,[70] a result supporting the simulation finding.

An interesting correlation between the torsional dynamics of the polymer backbone and the diffusion motion of hydration water can be detected by considering that these dynamical processes have the same activation energy. The conformational transition of dihedral angles formed by backbone carbon atoms involves a movement of side chain amide groups, surrounded by clusters of water molecules. The rearrangement of such clusters and the backbone torsions are therefore concerted and concur to the segmental dynamics of PNIPAM in the water soluble state.

We can then summarize the hydration features of PNIPAM in the undercooled solution on the basis of simulations results. At 243 and 253 K the polymer-water HBs and the hydrogen bonding connectivity of FHS water are the highest (Fig. 4 and Figures 5, 7, respectively). These elements indicate that the undercooled polymer solution is supported by both hydrophilic and hydrophobic hydration modalities, just like at the higher temperatures but below the coil-to-globule transition temperature. Overall, according to our simulations no UCST behaviour can be expected for the single chain of PNIPAM in water in the low temperatures domain. However, demixing in polymer aqueous solution can occur also without an intramolecular rearrangement but only with the inter-chain association and phase separation.[99] Typically these processes occur at higher polymer concentrations. Therefore, before ruling out the UCST of PNIPAM at low temperature from a simulation point of view, the atomistic MD simulation of solution models containing a consistent number of polymer chains should be performed. This is one of the perspectives of our study.

## Conclusions

This work explores the temperature dependence of structure and dynamics of water in the surrounding of PNIPAM in the diluted aqueous solution of this polymer. A particular care is taken in the realistic modeling of the polymer structure, according to the tacticity characteristics obtained in not stereo-selective syntheses.[40, 41] The investigated interval of temperature ranges from the undercooled solution to above the LCST of this polymer in water, i.e. in the range 243-323 K.[7] The polymer behaviour in the simulations reproduces the PNIPAM coil-to-globule transition at the expected temperature and the simulation results are in agreement with available experimental data on polymer hydration and dynamics. Our investigations add novel information and confirm previously proposed hypotheses on the features of the process of polymer de-hydration, which is concomitant with the chain rearrangement in the coil-to-globule transition: i) the amount of hydration water lost in the transition is relatively low, namely about 14 %; ii) the cooperativity of the transition is ascribable to the breaking of the HB network formed by water molecules in proximity of hydrophobic isopropyl groups; iii) the perturbation of the HB network formed between PNIPAM amide moieties and water is subsequent to the chain collapse; iv) the endothermal character of the coil-to-globule transition derives from a balance between breakage and formation of HBs in the surrounding of the polymer, where the predominant factor is the decrease of amide-water hydrogen bonds; v) across the transition temperature the water molecules surrounding PNIPAM hydrophobic groups experience a decreased structuring, as compared to bulk water. Moreover, the simulations highlight a correlation between PNIPAM segmental dynamics and water diffusion motion of bound water in the temperature interval corresponding to the thermodynamically stable aqueous solution, occurring with the same activation energy. Altogether these results show that PNIPAM is largely hydrated even above the LCST and that the coil-to globule transition is associated with a significant rearrangement of the hydration pattern.

This study for the first time also focuses on the PNIPAM aqueous solution at sub-zero temperature, with the aim to detect evidence of an UCST behaviour of this polymer in water at low temperatures. The answer provided by our simulations on this issue is negative, since hydration modality and water affinity at temperatures below the $T_{C \rightarrow G}$ do not show discontinuities. To clarify this point, future work is planned where simulations will be carried out at higher polymer concentrations and higher number of polymer chains. The presence of an UCST for PNIPAM aqueous solutions in the high temperatures domain, leading to a closed-loop mixing gap in the phase diagram, as observed for poly(ethylene oxide) in water, is predicted by the thermodynamic model of Rebelo et al.[100] However, at the expected UCST value, liquid water is not stable.[9] The occurrence of a high temperature UCST for PNIPAM in water was not yet experimentally observed. Nevertheless, within the analogy of PNIPAM with a protein system, a temperature induced globule-to-coil process, corresponding to the heat denaturation, could be expected. MD simulations at higher temperatures can contribute to clarify such issue.

Further perspectives of this study include the molecular description of PNIPAM microgels and nanogels at low temperatures, to selectively study the quenching of the dynamics of polymer and water in such systems.

## Conflicts of interest

There are no conflicts to declare.

## Acknowledgements

LT, EC and EZ acknowledge support from the European Research Council (ERC Consolidator Grant 681597, MIMIC). High performance computing resources and support from CINECA under the ISCRA initiative is gratefully acknowledged. Computing activity was carried out within the ISCRA C project HP10CYRFZZ.

## Notes and references

# Supplementary Information

**Table S1.** Stereochemistry and initial backbone conformation of PNIPAM 30-mer.

| Residue number | Configuration | Conformation |
|:---:|:---:|:---:|
| 1 | *l* | tt |
| 2 | *d* | tt |
| 3 | *l* | tt |
| 4 | *d* | g*t |
| 5 | *d* | g*t |
| 6 | *d* | g*t |
| 7 | *l* | gt |
| 8 | *l* | gt |
| 9 | *d* | tt |
| 10 | *l* | tt |
| 11 | *d* | tt |
| 12 | *l* | gt |
| 13 | *l* | gt |
| 14 | *l* | gt |
| 15 | *d* | tt |
| 16 | *l* | gt |
| 17 | *l* | gt |
| 18 | *l* | gt |
| 19 | *l* | gt |
| 20 | *l* | gt |
| 21 | *d* | g*t |
| 22 | *d* | g*t |
| 23 | *d* | g*t |
| 24 | *l* | gt |
| 25 | *l* | gt |
| 26 | *d* | g*t |
| 27 | *d* | g*t |
| 28 | *l* | tt |
| 29 | *d* | tt |
| 30 | *l* | - |

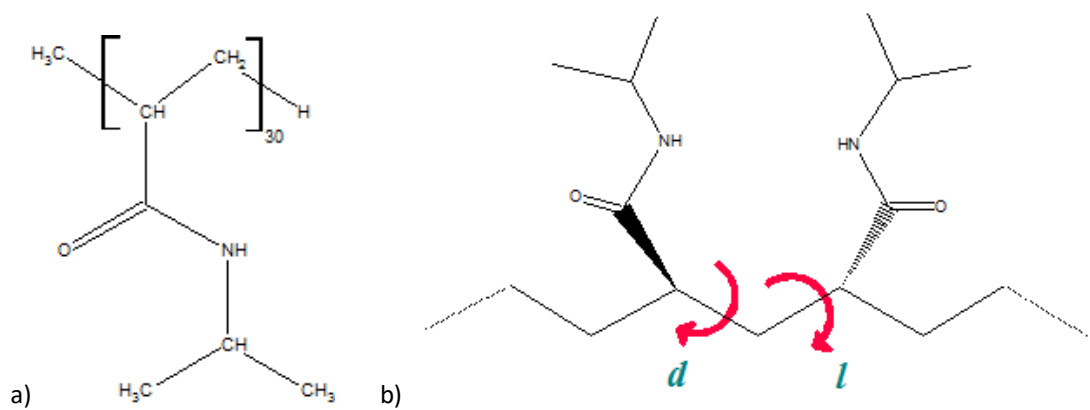

**Fig. S1** Illustration of the chemical structure of the PNIPAM 30-mer (a) and definition of the residues configuration (b). Arrows indicate dihedral backbones.

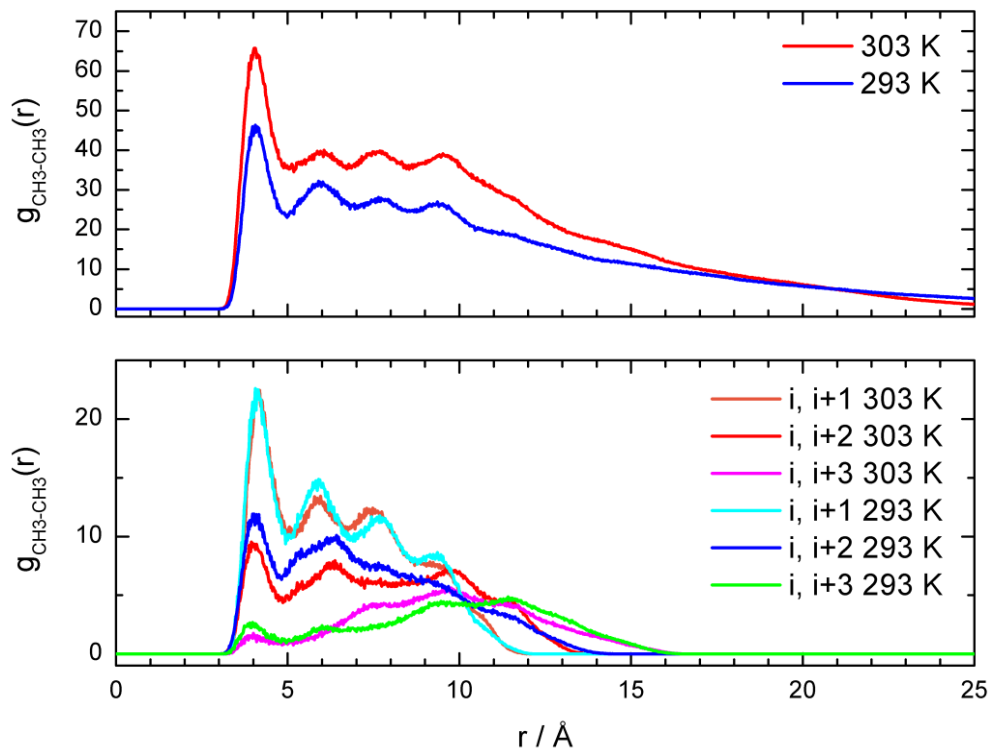

**Fig. S2** Pair distribution functions of PNIPAM methyl groups as calculated from the MD simulations. On the upper panel the total distributions at 303 K and 293 K are shown in blue and red, respectively. On the lower panel the contributions of the i, i+1 residues (orange at 303 K and light blue at 293 K), i, i+2 residues, (red at 303 K and blue at 293 K) and i, i+3 residues (magenta at 303 K and green at 293 K) are represented individually.

The clustering of methyl groups at 303 K, visible by comparing the radial distribution functions between methyl carbon atoms, $rdf_{CC}$, at 293 and 303 K (Fig. S2 of ESI), confirms that the backbone conformational transition observed at 303 K is concerted with a rearrangement of side chain groups to minimize the exposed hydrophobic surface area. Ahmed at al. proposed that the hydrophobic collapsed state may be formed by pushing the i, i + 3 isopropyl groups closer together to form local hydrophobic pockets.[51] We selectively analysed the $rdf_{CC}$ between pair of PNIPAM residues and found no increment of clustering between i, i + 3 isopropyl groups at 303 K.

**Table S2.** Average time of transitions of the backbone dihedrals $<\tau_b>$ and the dihedrals of the methyl groups $<\tau_m>$ as function of the temperature.

| Temperature (K) | $<\tau_b>$ (ns) | $<\tau_m>$ (ps) |
|:---:|:---:|:---:|
| 243 | -* | 340±20 |
| 253 | -* | 270±10 |
| 263 | 70±20 | 200±10 |
| 273 | 40±10 | 150±4 |
| 283 | 20±10 | 130±4 |
| 293 | 10±2 | 110±2 |
| 303 | 10±5 | 90±2 |
| 313 | 5±1 | 74±2 |
| 323 | 5±1 | 60±0.3 |

*Average time of transition not defined due to reduced sampling.*

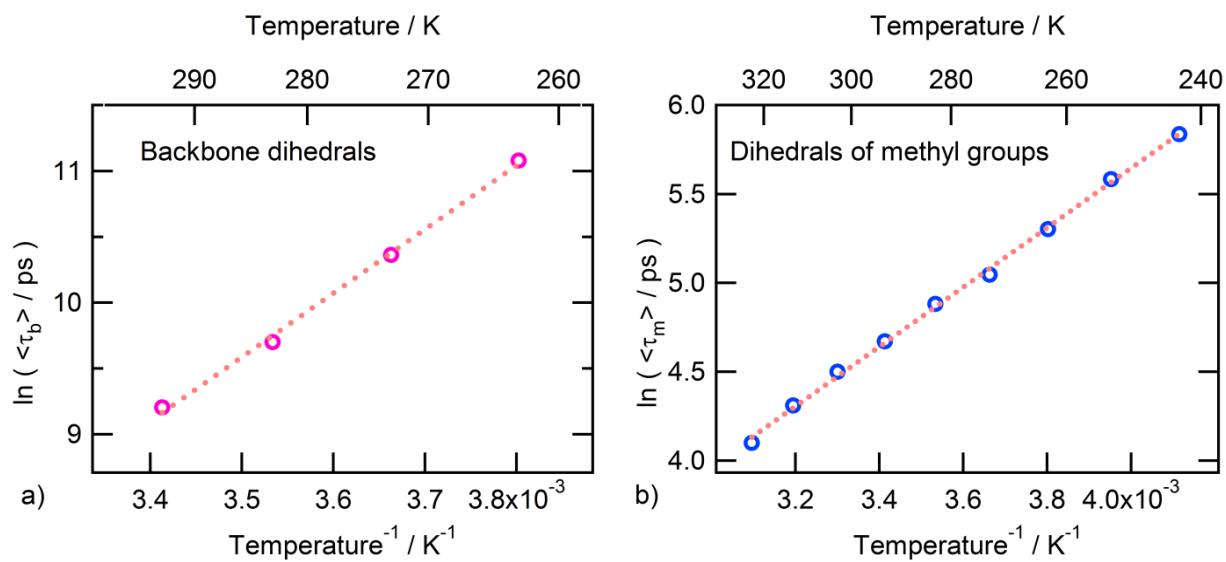

**Fig. S3** Arrhenius plot of the average lifetime of a rotational state of a backbone dihedral (a) and of a dihedral of a methyl group (b). Dashed line shows the linear fit of the data.

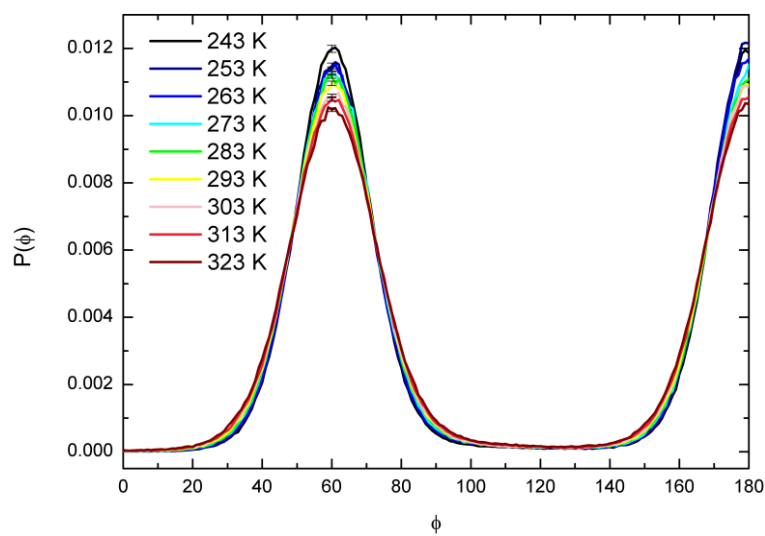

**Fig. S4** Distribution of dihedral angles of PNIPAM methyl groups as a function of the temperature.

**Table S3.** Average number of hydration water molecules per PNIPAM residue.

| Temperature (K) | Hydration water molecules *per* residue |
|:---:|:---:|
| 243 | 14.3±0.3 |
| 253 | 14.5±0.3 |
| 263 | 13.9±0.3 |
| 273 | 14.0±0.4 |
| 283 | 14.2±0.4 |
| 293 | 13.7±0.5 |
| 303 | 11.5±0.5 |
| 313 | 12.2±0.6 |
| 323 | 11.4±0.6 |

*Time averaged values and standard deviation.*

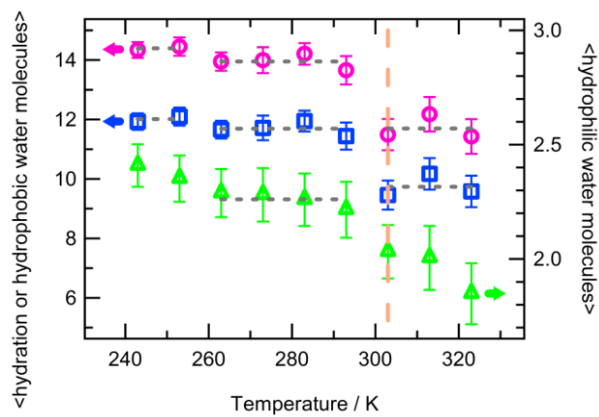

**Fig. S5** Temperature dependence of the average number of FHS water molecules *per* residue classified as total hydration water molecules (pink circles), hydrophobic water molecules (blue squares) and hydrophilic water molecules (green triangles). Data represent time averaged values and standard deviation. Dashed lines are guide to the eye.

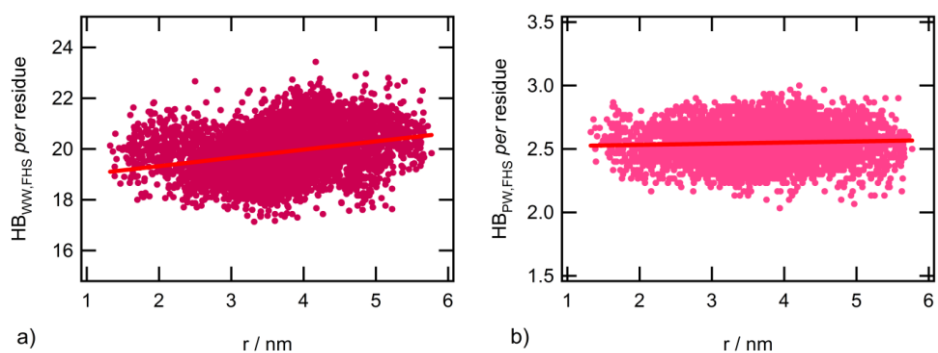

a)                                                          b)

**Fig. S6** a) Correlation between hydrogen bonding of water molecules *per* repeating unit in the FHS and end to end distance (r), for PNIPAM 30-mer at 293 K. b) Correlation between PNIPAM-water hydrogen bonding and end to end distance, for PNIPAM 30-mer at 293 K.

**Table S4.** Comparison between hydrogen bonding of bulk water and of water hydrating PNIPAM hydrophobic groups.*

| Temperature (K) | 293 | 303 | 313 |
|---|---|---|---|
| Number of HBs (*per* water molecule) in bulk water | $2.18 \pm 0.01$ | $2.20 \pm 0.01$ | $2.20 \pm 0.01$ |
| Number of HBs (*per* water molecule) in the first hydration shell of PNIPAM isopropyl groups | $1.42 \pm 0.03$ | $1.40 \pm 0.03$ | $1.40 \pm 0.03$ |

*Errors are estimated by standard deviation.

**Supplementary movie**

Time evolution of PNIPAM chain conformation and its first hydration shell at 283 K (on the left) and 303 K (on the right) over the first 150 ns of simulation. PNIPAM backbone atoms are shown in white, hydrophilic water oxygen atoms in red and hydrophobic water oxygen atoms in blue.